\documentclass[pdflatex]{sn-jnl}

\usepackage{graphicx}%
\usepackage{multirow}%
\usepackage{amssymb}
\usepackage{amsmath,amssymb,amsfonts}%
\usepackage{amsthm}%
\usepackage{mathrsfs}%
\usepackage[title]{appendix}%
\usepackage{xcolor}%
\usepackage{textcomp}%
\usepackage{manyfoot}%
\usepackage{booktabs}%
\usepackage{algorithm}%
\usepackage{algorithmicx}%
\usepackage{algpseudocode}%
\usepackage{listings}%
\usepackage[normalem]{ulem}
\usepackage[
backend=biber,
style=nejm,
sorting=none,
isbn=false,
url=false,
eprint=false,
date=year
]{biblatex} 
\begin{document}

\title{Giant thermal modulation via a semiconductor-superconductor photonic field-effect heat transistor}
\author*[1]{\fnm{Sebastiano} \sur{Battisti}}\email{sebastiano.battisti@sns.it}
\author[1]{\fnm{Matteo} \sur{Pioldi}}
\author[1]{\fnm{Alessandro} \sur{Paghi}}
\author[1]{\fnm{Giorgio} \sur{De Simoni}}
\author[1]{\fnm{Alessandro} \sur{Braggio}}
\author[1]{\fnm{Giulio} \sur{Senesi}}
\author[1]{\fnm{Lucia} \sur{Sorba}}
\author*[1]{\fnm{Francesco} \sur{Giazotto}}\email{francesco.giazotto@sns.it}

\affil[1]{\orgdiv{NEST}, \orgname{Istituto Nanoscienze-CNR and Scuola Normale Superiore}, \orgaddress{\city{Pisa}, \postcode{56127}, \country{Italy}}}

\abstract{
We present a groundbreaking demonstration of thermal modulation in a field-effect-controllable semiconductor-superconductor hybrid structure, wherein the heating mechanism is exclusively radiative. The architecture comprises two reservoirs separated by $\sim 1$ mm and interconnected via a completely non-galvanic electrical circuit, enabling the transfer of black-body radiation from the hot to the cold reservoir. Our device utilizes a superconducting Josephson field-effect transistor to achieve magnetic-field-free gate-tunable regulation of heat currents within the circuit. 
While prior studies have indicated the potential for electrostatic modulation of thermal transport properties, our framework demonstrates a temperature modulation of up to $\sim 45$ mK, exceeding prior findings by more than an order of magnitude. Furthermore, it proves a thermal transimpedance of $\sim 20$ mK/V at a bath temperature of $30$ mK. The development of such systems holds substantial promise for advancing heat management and routing in quantum chips and radiation sensors, as it enables precise nonlocal control of heat flow towards a designated structure, even when the heat source is distant and non-galvanically coupled.
}
\maketitle

\section*{Introduction}
In recent decades, the field of photonic heat transport within superconducting circuits at cryogenic temperatures has garnered substantial academic interest \cite{maillet2020,subero2023,meschke2006,ronzani2018,gubaydullin2022,partanen2016,thomas2019,pekola2021}. This phenomenon arises from black-body radiation mediated by Johnson-Nyquist noise, which occurs when two resistive reservoirs at different temperatures are interconnected through an electrical circuit \cite{johnson1928,nyquist1928}.
Incorporating a tunable element within the coupling circuit enables modulation of photonic heat exchange between the two reservoirs. This concept has been examined in several theoretical \cite{ojanen2008,marchegiani2021,paolucci2017b,tam2023,yeyati2024} as well as experimental studies, using both external magnetic \cite{subero2023,meschke2006} and electrostatic fields \cite{maillet2020}. Even the application of photonic thermal transport to qubit structures was explored \cite{gubaydullin2022}, using external magnetic fields to modulate the coupling.

The performance of these devices is typically measured by the temperature modulation achievable in one of the reservoirs. State-of-the-art devices that employ magnetic fields can achieve temperature modulations as high as $\delta T\simeq 5$ mK \cite{meschke2006,subero2023}. In these structures, the manipulation of photonic heat flow is usually achieved by introducing a superconducting quantum interference device (SQUID), whose impedance is purely inductive and depends on the applied magnetic field \cite{squidhandbook2006}. Indeed, the magnetic flux that pierces the SQUID loop alters the critical current of the whole ring, hence varying its Josephson inductance periodically \cite{squidhandbook2006}. Eventually, the heat flow modulations of those devices are periodic in the magnetic field, resembling the periodic impedance variation of a SQUID loop. Avoiding the use of magnetic fields can significantly improve the usability and implementation of photonic heat transport in superconducting quantum technologies. Complete electrostatic control of photonic heat transport was demonstrated using gated superconducting single-electron transistors (SETs) \cite{maillet2020}. These structures also showed periodic modulation of heat flow due to the periodic impedance behavior of a superconducting SET as a function of the applied electrical field \cite{averin1991}. The performance in terms of temperature modulations of these devices was $\delta T \simeq2$ mK.

In this study, we investigate the performance of a hybrid semiconductor-superconductor photonic thermal device that electrostatically tunes thermal transport. This device is implemented in an unintentionally n-type doped InAs layer, which is grown on top of an insulating InAlAs metamorphic buffer, commonly referred to as InAs-on-Insulator (InAsOI) \cite{paghi2025c,battisti2024}, due to its similarity to the prevalent Silicon-on-Insulator (SOI) technology.
The objective of our research is to achieve tunable heat exchange within a photonic transport circuit by deploying a scalable, straightforward top-gate architecture, thereby circumventing the need for magnetic controls or more complex structures for heat flow manipulation. The adoption of InAsOI presents several advantages. Primarily, its semiconductive nature permits the field-effect regulation of its conduction properties. Additionally, InAs has been demonstrated to be effective in achieving superconductor-semiconductor-superconductor (SSmS) Josephson junctions (JJs), which can be adjusted via a top-gate control electrode that functions as a Josephson field-effect transistor (JoFET) \cite{paghi2025b}.
An additional and substantial merit of our proposal lies in the demonstrably weak electron-phonon (e-ph) coupling, as empirically demonstrated \cite{battisti2024}, which positions it as an exemplary candidate for applications in caloritronics.

Here, we investigate two resistive reservoirs separated by a distance of $\sim1$ mm to demonstrate the robustness of this platform at macroscopic scales and to avoid unintended phonon-mediated thermal coupling through the substrate. We connect two terminals via a non-galvanic circuit that uses a JoFET to regulate photonic heat transfer. Our findings on temperature modulation exceed the current state of the art by more than an order of magnitude \cite{maillet2020}. This device represents a substantial advancement in areas such as coherent caloritronics \cite{fornieri2017,martinez-perez2014,fornieri2016b,martinez-perez2013,hwang2020,giazottoOpportunities2006}, thermal and energy manipulation \cite{fornieri2016,martinez-perez2015,fornieri2014}, bolometric radiation detection \cite{virtanen2018,kokkoniemi2019,kokkoniemi2020}, thermal logic gates and memory systems \cite{wang2007,wang2008,li2012,paolucci2018,guarcello2018,ligato2022}, and for the whole field of superconducting quantum technologies \cite{dowling2003}.

\subsection*{Conceptualization of the Superconducting Photonic Field-Effect Heat Transistor}
\begin{figure*}[t!]
  \includegraphics[width=1\columnwidth, trim= 0cm 0cm 0cm 0cm]{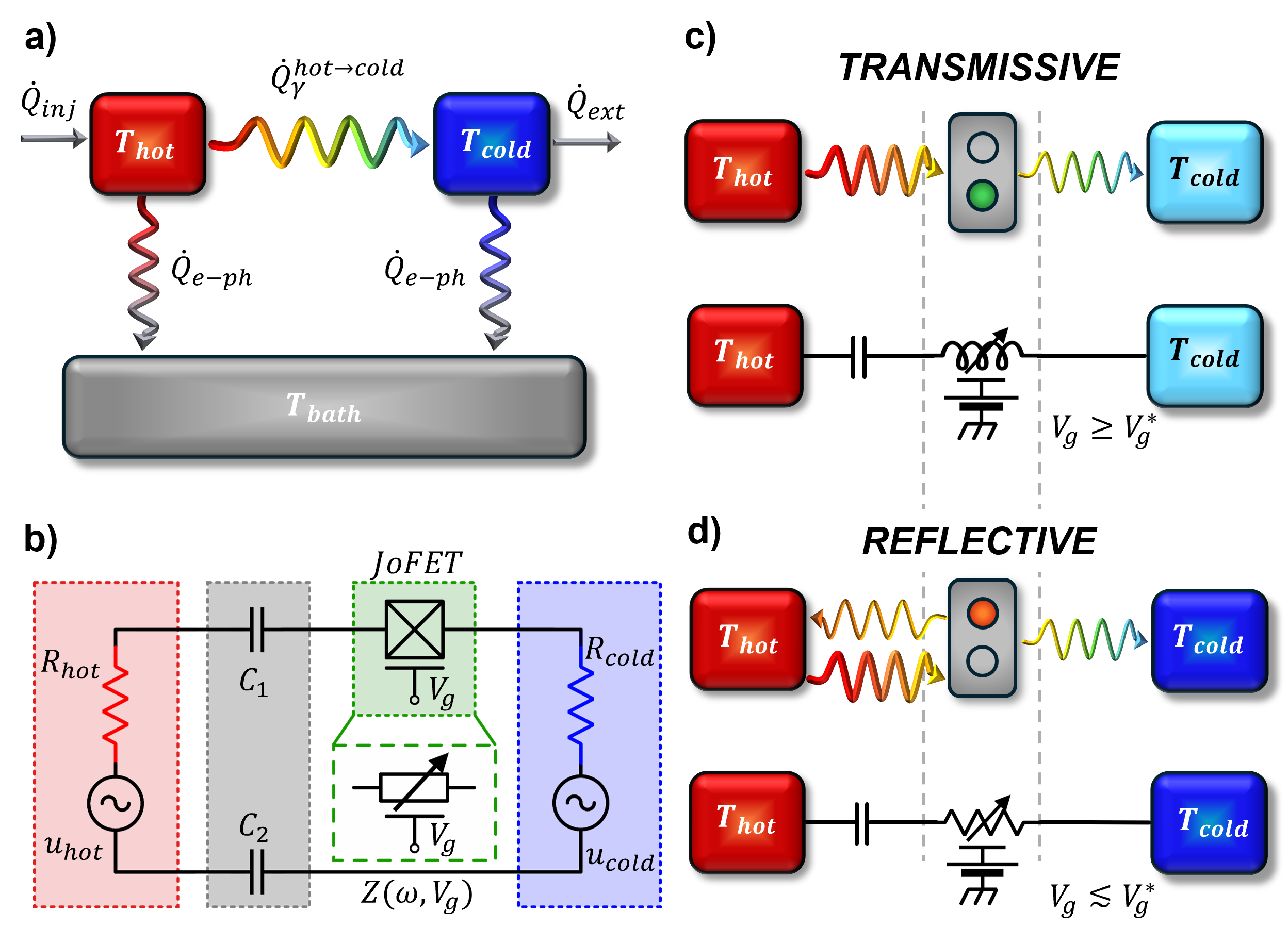}
    \caption{\textbf{Schematic representation of the experiment.} \textbf{a}, Heat balance scheme of the system: $\dot{Q}_{inj}$ represents the injected power into the hot resistor, $\dot{Q}_{e-ph}$ denotes the reservoir electron-phonon heat exchanges towards the thermal bath, $\dot{Q}_\gamma$ indicates the photonic heat current, and $\dot{Q}_{ext}$ refers to other parasitic heat currents losses in the cold resistor. \textbf{b}, Lumped element model of the circuit. The hot/cold parts of the circuit (red and blue boxes) are represented with resistive reservoirs $R_{hot}$/$R_{cold}$ and effective Johnson-Nyquist AC voltage sources $u_{hot}$/$u_{cold}$. The circuit comprises the decoupling capacitors $C_{1,2}$ (grey box) and the JoFET, which is controllable by the gate voltage $V_g$ (green box). The JoFET is schematized as a frequency- and voltage-dependent impedance $Z(\omega,V_g)$. \textbf{c}-\textbf{d}, Functioning principle of the experiment. The two reservoirs are interconnected by a circuit whose total complex impedance can be adjusted via an external gate voltage $V_g$. When the JoFET mediates supercurrent (\textbf{c}), the impedance is purely \textit{inductive}, allowing for maximum photonic heat transfer through the system (green semaphore). Conversely, when the JoFET is in the \textit{resistive} state (\textbf{d}), it creates a strong impedance mismatch between the two reservoirs, thereby increasing the back-reflection of the radiative heat flow (red semaphore).}
  \label{Fig1}
\end{figure*}

The thermal model of our experiment is shown in Fig. \ref{Fig1} (a). Two different resistive hot and cold reservoirs are maintained at different temperatures $T_{hot}$ and $T_{cold}$, respectively. They are thermally coupled to the phonon bath, which is assumed to be maintained at a constant common temperature $T_{bath}$. This assumption comes from the approximation of the vanishing Kapitza resistance between the phonons of the resistors and those of the substrate \cite{giazottoOpportunities2006}. $\dot{Q}_{inj}$ represents the power injected into the hot reservoir, while $\dot{Q}_{ext}$ refers to the incoming/outgoing parasitic heat currents in the cold reservoir. 

The $\dot{Q}_{e-ph}$ denotes the electron-phonon (e-ph) heat exchange from the hot and cold reservoirs to the phonon bath.
The hot and cold reservoirs are themselves part of the electrical circuit, as illustrated in Fig. \ref{Fig1} (b). They can be described in the lumped element approximation as resistors $R_{hot}$ and $R_{cold}$ and, correspondingly, as two virtual Johnson-Nyquist thermal noise voltage sources $u_{hot}$ and $u_{cold}$ \cite{johnson1928,nyquist1928}. Indeed, thermal fluctuations in the reservoirs induce voltage fluctuations across them, reflecting the exchange of black-body radiation between them. 
However, since the typical thermal photon wavelength is $\lambda_\gamma=hc/(n_{InAs}k_BT)$ ($h$ the Planck constant, $c$ the speed of light, $k_B T$ the Boltzmann thermal energy, and $n_{InAs}\sim 4$ is the refractive index of InAs \cite{aspnes1983}) one finds that 
at very low temperatures $T\sim100$ mK, the thermal photon wavelength is $\lambda_\gamma\sim 3$ cm on a macroscopic scale. This is significantly longer than the characteristic size of the electrical elements, so this thoroughly explains the lumped element approximation that is adopted hereafter \cite{maillet2020}. 

Furthermore, the hot and cold parts are galvanically decoupled thanks to capacitors $C_1$ and $C_2$, enabling independent low-frequency DC networks.
In this framework, the modulation of photonic heat conduction, which enables thermal exchange, can be achieved by introducing a tunable frequency-dependent impedance between the two resistive reservoirs.

In this context, a gate-controlled dissipationless Josephson junction, i.e., a JoFET, is an ideal candidate to achieve this task. Indeed, the gate voltage $V_g$ allows adjustment of the electrical impedance $Z(\omega,V_g)$ from purely reactive to purely resistive. 
$Z(\omega,V_g)$ exhibits two primary regimes: it can act as a Josephson inductance $L_J(V_g)$ or as resistance $R_J(V_g)$. 
In fact, when $V_g$ is greater than a certain threshold $V_g^*$, JoFET mediates a supercurrent and effectively has a negligible resistance, presenting primarily an inductive component. At the same time, for $V_g\lesssim V_g^*$, the JoFET is no longer proximitized, acting primarily as a dissipative element.\\

The tuning of the JoFET impedance modulates the flux of heat because, in such a limit, the photonic thermal current ($\dot{Q}^{i\rightarrow k}_\gamma$)  between two generic reservoirs (see Fig. \ref{Fig1} (a)) can be calculated using the scattering formula \cite{pascal2011}
\begin{equation}  
\dot{Q}^{i \rightarrow k}_\gamma=\int_0^\infty \tau_{ik}(\omega)\hbar\omega[n_{i}(\omega)-n_{k}(\omega)]\frac{d\omega}{2\pi},
    \vspace{5mm}
\end{equation}
where $n_{j}(\omega)=1/[\exp(\hbar\omega/k_BT_{j})-1]$ is the Bose-Einstein distribution for the  photon population in the $j$th-reservoir, $\hbar=h/2\pi$, and $\omega$ is the angular frequency. The power transmission coefficient is obtained as $\tau_{ik}(\omega,V_g)=\frac{4R_{i}R_{k}}{|Z_{tot}(\omega,V_g)|^2}$, where $Z_{tot}(\omega,V_g)$ is the total series impedance of the circuit and $R_{i,k}$ are the reservoir resistances \cite{pascal2011}.
Under the assumption of $C_1=C_2=C$, we can approximate the power transmission coefficient $\tau_{ik}(\omega,V_g)$ as

\vspace{4mm}
\begin{equation}
\tau_{ik}(\omega,V_g)\approx
    \begin{cases}
    \frac{4R_iR_k}{[R_{hot}+R_{cold}]^2+[\omega L_J(V_g)-\frac{4}{\omega C}]^2} \;\;\; \text{for} \;\;\; V_g\geq V_g^*\\
    \\
    \frac{4R_iR_k}{[R_{hot}+R_{cold}+R_J(V_g)]^2+[\frac{4}{\omega C}]^2} \;\;\; \text{for} \;\;\; V_g \lesssim V_g^*
    \end{cases}
    \vspace{5mm}
\end{equation}

The expected behavior of the proposed setup is illustrated in Figs. \ref{Fig1} (c,d), where two distinct regimes are depicted for $\dot{Q}^{hot \rightarrow cold}_\gamma$. The case of the transmissive photonic channel (Fig. \ref{Fig1} (c)) is realized when the JoFET shows an inductive behavior: the impedance matching of the coupling circuit is optimal, and the heat transfer due to the photonic current is maximum. The coupling circuit is conceptually depicted as a green semaphore in this case. By contrast, the reflective photonic channel case (Fig. \ref{Fig1} (d)) is realized when the JoFET acts as a resistor, resulting in a degraded impedance matching and a reduction in the net radiative heat transfer through the coupling circuit. The coupling circuit is therefore conceptually depicted as a red semaphore.

\subsection*{Fabrication Process and Experimental Setup}

\begin{figure*}[t!]
  \includegraphics[width=0.9\columnwidth, trim= 0cm 0cm 2cm 0cm]{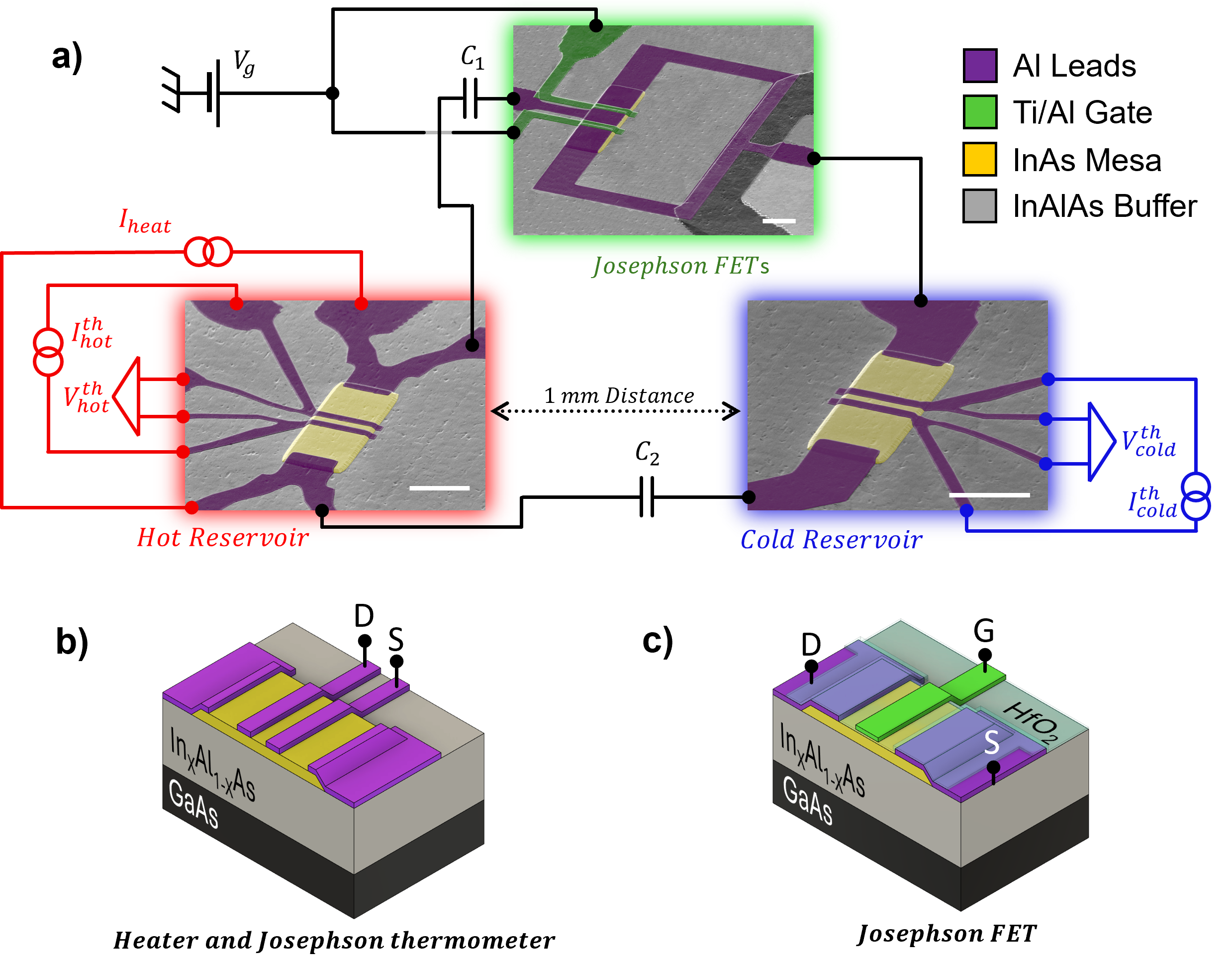}
    \caption{\textbf{Structure and wiring of the gate-tunable photonic heat transfer experiment.} \textbf{a}, Schematic circuit including SEM photographs (the white scale bars in those photos are $5$ $\mu$m) of the devices integrated on the InAsOI platform, wiring, and supply/measurement setup. Red- and blue-shaded SEM images show the hot and cold resistors, respectively, equipped with JJ thermometers. The green-shaded SEM image shows the gate-tunable photonic heat modulator, achieved by using two JoFETs connected in parallel and driven simultaneously. The ring-like configuration realizes a superconducting interferometer (SQUID). The hot and cold resistors are non-galvanically connected via the two capacitors. Red and blue wiring and schematics refer to the supply/measurement setups of the hot and cold reservoirs, respectively. \textbf{b}, 3D representation of the hot and cold resistors equipped with the Josephson thermometer. The two Al fingers in the middle define the SSmS JJ used to measure the electronic temperature. At the same time, the lateral Al leads connect the InAs mesa to the rest of the circuit and also provide the heating current. \textbf{c}, 3D representation of a single JoFET. The superconducting source and drain leads on the side define the SSmS junction, with the InAs mesa patterned on the electrical insulating substrate. The gate finger in the middle is non-galvanically connected to the InAs channel via the HfO$_2$ insulating layer. It is used to tune the electron density of the InAs channel, thereby affecting the photonic heat transfer. }
  \label{Fig2}
\end{figure*}
The circuit fabrication process was developed on the InAsOI platform, a Molecular Beam Epitaxy-grown heterostructure consisting, from bottom to top, of a $500$ $\mu$m-thick semi-insulating GaAs (100) substrate, a $1.3$ $\mu$m-thick step-graded In$_X$Al$_{1-X}$ as metamorphic buffer with $X$ increasing from $0.15$ to $0.81$, and a $100$ nm-thick InAs semiconductive epilayer \cite{arif2024,benali2022,senesi2025}.

The InAs layer is unintentionally n-doped, exhibiting a sheet electron density ($n_{2D}$) of $1.20\times 10^{12}$ cm$^{-2}$. The metamorphic InAlAs buffer behaves as an electrical insulator at cryogenic temperatures.
Active and passive devices were fabricated using e-beam evaporation of superconducting Al, atomic layer deposition (ALD) of HfO$_2$ insulator, e-beam lithography, and Al and III-V semiconductor wet etching steps, followed by a lift-off process. More details on the fabrication process can be found in the \textit{Methods} section.
 InAsOI interfaces very well with superconductors due to the lack of a Schottky barrier at contact with the metal, allowing us to create SSmS JJs and JoFETs \cite{paghi2025c,paghi2025b}.

Figure \ref{Fig2} (a) illustrates the structure and wiring of the fabricated circuit, accompanied by scanning electron microscopy photographs of the fabricated devices. The chip employs three InAs mesas, highlighted in yellow. The InAs mesas were defined via selective wet etching of the excess InAs epilayer and electrically decoupled by the electrical insulating behavior of the InAlAs metamorphic buffer at cryogenic temperatures, here depicted in gray. 
The superconducting Al leads, along with the signal lines, are delineated in violet, whereas the Ti/Al gate lines are distinguished by green coloring. Gate lines are electrically isolated from signal traces by a 31-nm-thick HfO$_2$ insulating layer deposited by ALD.

The InAs mesas were configured in two different arrangements: hot and cold resistive thermal reservoirs equipped with heaters and Josephson thermometers (red and blue photographs in Fig. \ref{Fig2} (a), schematized in Fig. \ref{Fig2} (b)) and a voltage-controlled JoFET used as a thermal current modulator (green photograph in Fig. \ref{Fig2} (a), schematized in Fig. \ref{Fig2} (c)).
The resistive reservoirs ($10$ $\mu$m long, $5$ $\mu$m wide) are macroscopically spaced 1 mm 
apart on the chip. The value of their resistances was measured to be $R_{hot}=R_{cold}=360\pm10$ $\Omega$. Both reservoirs were equipped with Josephson thermometers (violet source (S) and drain (D) contacts in Fig. \ref{Fig2} (b)) for the purpose of monitoring their electronic temperature, evaluated by measuring the temperature-sensitive switching supercurrent of the SSmS JJs defined between the Al superconducting fingers positioned in the middle of the InAs mesas. The inter-finger distances defining the JJ lengths are $450$ nm and $550$ nm for the hot and cold reservoirs, respectively, to get different sensitivity ranges in the electronic temperature (see SI for thermometry calibration details). 

The temperature measurement is performed by supplying a current $I^{th}_{hot,cold}$ to drive the JJ junction, while the voltage drop across it $V^{th}_{hot,cold}$ was measured to identify the switching current and convert it by calibration to the JJ electron temperatures (see SI for thermometry calibration). 
It is essential to note that the temperature measurements on the mesas were conducted non-simultaneously to mitigate potential thermal disturbances to the system arising from switching Josephson thermometers.

The external control of the electronic temperature of the heated hot resistor is obtained by providing an additional external current $I_{heat}$ directly injected from two additional Al lines, inducing Joule heating $\dot{Q}_{inj}=R_{hot}I^2_{heat}$ directly within the InAs mesa. This current is supplied by an externally floating current generator that is electrically isolated from other sources to prevent the occurrence of unintended ground loops. 

Incorporation of components $C_1$ and $C_2$ is essential to achieve direct current (DC) isolation between the heating circuit and the cold reservoir. Indeed, they ensure galvanic isolation between the two reservoirs. These components have been strategically integrated between the hot and cold resistive reservoirs, as well as between the hot resistive reservoir and the JoFET, to complete the circuit configuration. Each capacitor consists of a series configuration of two identical metal-insulator-metal parallel-plate capacitors, with Al electrodes (area of $130$ $\mu$m$^2$) separated by the HfO$_2$ insulator, resulting in an inter-electrode distance of 31 nm. The design anticipates a capacitance $C_1=C_2=C$ of $\simeq 70$ pF \cite{paghi2025}.

In contrast to the simplified scheme illustrated in Fig. \ref{Fig1} (b), in the actual setup, the voltage-tunable electrical impedance used as the thermal transport modulator is realized using two nominally identical JoFETs implementing an SSmS superconducting quantum interference device (SQUID) equipped with a superconducting flux line.
However, it was determined that the magnetic field-induced inductance variation exhibited an insignificant effect on the transmission coefficient $\tau_{ik}$ (refer to the discussion in the SI) as a result of the excessively large screening parameter of the SQUID. Consequently, a common gate voltage $V_g$ was applied to both JoFETs, which operate practically as a single JoFET element, having the sole mechanism of regulating photonic heat transport.
The adopted InAs mesa is $15$ $\mu$m long, $5$ $\mu$m wide, with a $4$ $\mu$m separation between the two JoFETs. The interelectrode distance of the gate-tuned JJs is $\sim450$ nm. The Ti/Al gate fingers of $1.5$ $\mu$m in width are centered on the fabricated JJs and separated from the SSmS channel by the HfO$_2$ layer described above. Details on the electrical properties of the JoFETs are reported in the following section.

\subsection*{Josephson Field Effect Transistor Electrical Characterization}

\begin{figure*}[t!]
  \includegraphics[width=1.1\columnwidth, trim= 0cm 1cm 0cm 0cm]{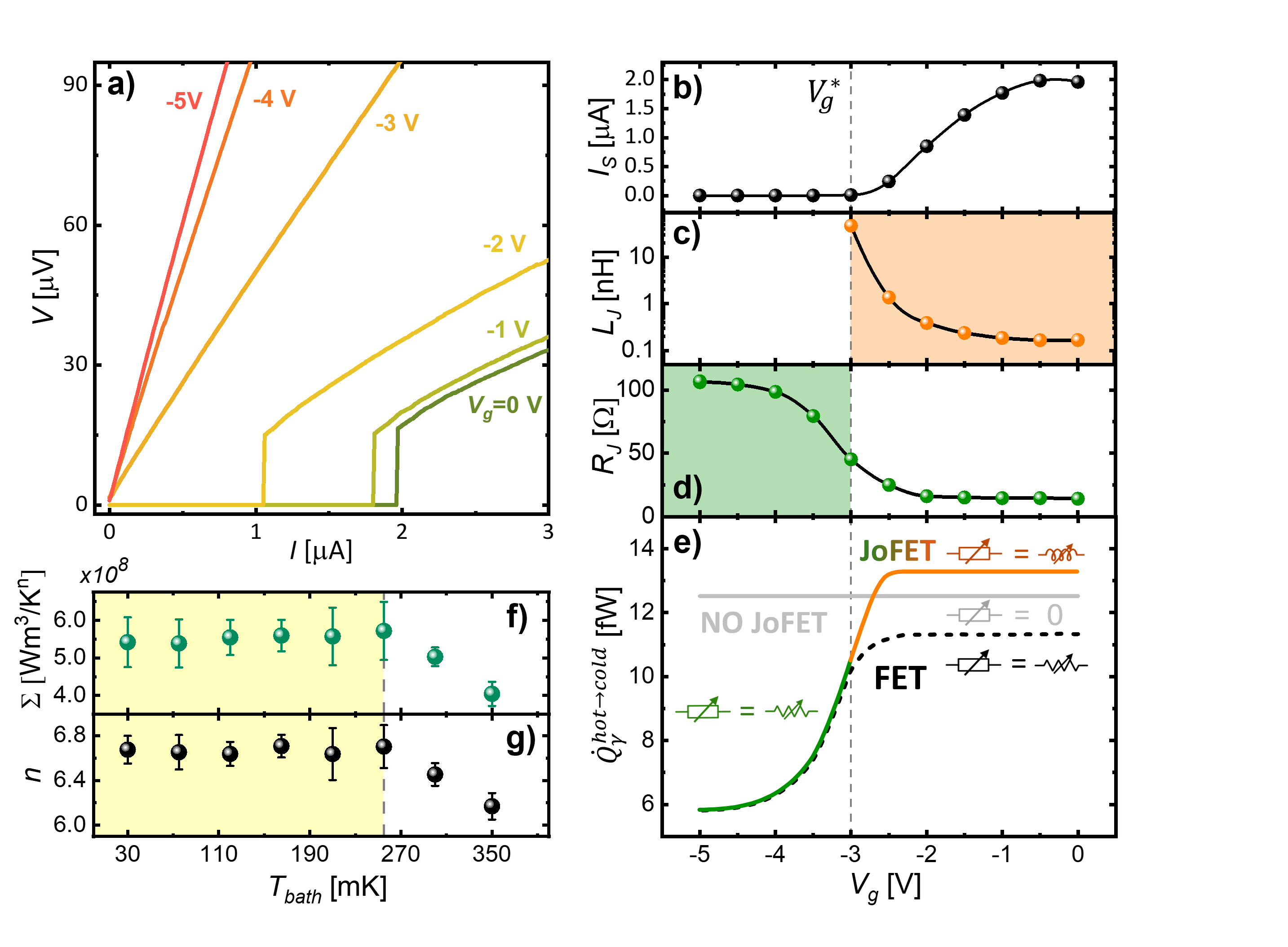}
    \caption{\textbf{Electrical characterization of the Josephson field effect transistor.} \textbf{a}, Gate-dependent $V-I$ characteristics of the JoFET. \textbf{b}, JoFET gate-dependent switching current. \textbf{c}, JoFET linear Josephson inductance plotted in logarithmic scale. \textbf{d}, JoFET normal-state resistance. Orange and green shaded boxes identify the dissipationless and resistive JoFET regimes, respectively. \textbf{e}, Numerical calculation of the gate-dependent photonic heat current from the hot to the cold reservoir. The orange and green curves refer to the JoFET in the inductive (orange solid line) and resistive (green solid line) regimes. The black-dashed curve represents a conventional non-superconducting FET acting as a gate-tunable resistor with value (d); the gray curve represents the case with no photon heat modulator. \textbf{f}-\textbf{g}, Experimental determination of the electron-phonon interaction parameters. (f) reports the e-ph coupling constant ($\Sigma$) while (g) depicts the power law constant ($n$)    as a function of $T_{bath}$. The yellow shaded region indicates the validity range of our analysis.}
  \label{Fig3}
\end{figure*}
The JoFET was first characterized in a dilution refrigerator equipped with filtered DC lines (more details of the measurement setup are provided in the \textit{Methods} section), where electrical characterization was performed at $30$ mK. 
Figure \ref{Fig3} (a) presents gate-dependent V-I curves of a twin JoFET structure present on the same chip, considered representative of that integrated in the electrical circuit. The switching current $I_S$ was extracted from the V-I curves as the largest current at 0-voltage drop, while the normal state resistance $R_J$ is estimated with a linear fit of the slope of the V-I curve well above $I_S$. By decreasing $V_g$, a reduction in the switching current is observed, as well as an increase in the slope of the V-I curve. This is attributed to the n-type behavior of the InAsOI-based JoFET, in which a reduction in the gate voltage reduces the density of the majority carriers (electrons), thereby suppressing Andreev transport. This is in complete agreement with our previous findings \cite{paghi2025c,paghi2025b}.

Figure \ref{Fig3} (b) reports the gate dependence of $I_S$. 
At a voltage of $V_g\lesssim V_g^*=-3$ V, the switching current reaches a value of zero, thereby eliminating the non-dissipative characteristics of the JoFET. Assuming that the JoFET operates in the linear response regime, its Josephson inductance in the superconducting state can be evaluated using the relation $L_J(V_g)=\hbar/2e I_s(V_g)$ ($e$ is the elementary charge) and plotted as a function of $V_g$ in Fig. \ref{Fig3} (c) on a logarithmic scale. Inductance $L_J$ increases with decreasing $V_g$, in agreement with the suppression of the switching current, eventually diverging to $+\infty$ for $V_g\to V_g^*$ since $I_S\to0$.

The normal-state resistance $R_J$ of JoFET also represents an essential parameter in our analysis, with the curve $R_J$ vs. $V_g$ depicted in Fig. \ref{Fig3} (d). Unlike $I_S$, $R_J$ increases by reducing the gate voltage while showing a sigmoidal trend.

The two shaded colored regions depicted in Fig. \ref{Fig3} (c,d) represent the two JoFET regimes: mainly inductive ($V_g\geq V_g^*$) and resistive ($V_g \lesssim  V_g^*$) operating ranges, identified in orange and green, respectively. 
The delimitation between these regions is established at $V_g^*$, the critical voltage threshold at which the JoFET completely transitions from a supercurrent-dominated inductive state to a resistive state.

To gain a more comprehensive understanding of the gate-dependent photonic heat transfer tunability of JoFET, Fig. \ref{Fig3} (f) reports the gate-dependent value of $\dot{Q}_\gamma^{hot \rightarrow cold}$, numerically evaluated in the presence of JoFET embedded in the circuit reported in Fig. \ref{Fig1} (b). This analysis utilizes Eqs. (1) and (2) while maintaining a constant temperature in the two reservoirs (i.e., $T_{hot}=350$ mK and $T_{cold}=30$ mK).
The orange part of the depicted curve indicates the behavior of the heat current when the JoFET is in the superconducting inductive state for $V_g\geq V_g^*$. In contrast, the green part corresponds to the resistive-dominated regime observed when $V_g \lesssim  V_g^*$. The transport of photonic heat exhibits a monotonic decrease in response to the decline in gate voltage, with a pronounced reduction observed at $V_g\lesssim -2.5$ V. This is related to the significant increase in the $L_J$ value in the range $V_g^*$$\leq V_g<-2.5$ V and, eventually, in the transition to the resistive regime for $V_g\lesssim V^*_g$. This numerical analysis elucidates the two regimes shown in Fig. \ref{Fig1} (c,d), where the photonic \textit{transmissive} regime corresponds to maximum heat transfer, and the photonic \textit{reflective} regime corresponds to minimal heat transfer.

The advantage of the integration of a superconducting JoFET, which can switch between inductive and resistive responses, is made evident by the comparative analysis conducted with a non-superconducting FET (black dashed line in Fig. \ref{Fig3} (e)) acting as a gate-tunable resistor, characterized by the resistive behavior reported in Fig. \ref{Fig3} (d) alone. The comparison reveals that, for $V_g \lesssim V^*_g$, both configurations produce identical results. However, for $V_g\geq V^*_g$, the inductive behavior of JoFET results in an increased heat transfer rate, thereby providing a more pronounced contrast between the photonic transmissive and reflective regimes.

To facilitate a thorough analysis, we have examined the scenario in which modulating elements are absent from the circuit depicted in Fig. \ref{Fig1} (b) (represented by the gray curve in Fig. \ref{Fig3} (e)).
This curve does not show a dependence on the gate voltage; however, it reveals an intriguing attribute of our configuration. Specifically, heat transfer within the photonic transmissive configuration plateau is enhanced in the presence of the JoFET. This phenomenon is attributed to the superior impedance matching throughout the circuit when an inductor is integrated. This can be analytically demonstrated by inspecting the denominator of $\tau_{ik}(\omega,V_g)$ in Eq. 2 for $V_g > -2.5$ V. The presence of an inductive (reactive) component, partially compensating for the capacitive terms, decreases the imaginary component of the total impedance $Z_{tot}(\omega,V_g)$ at the denominator, consequently increasing the value of $\tau_{ik}$. 
Hence, it can be inferred that integrating a JoFET results in more pronounced temperature variations than a traditional FET and provides superior transparency in the transmissive configuration compared to the case without any tuning component. This unique technique of gate-tunable impedance matching exhibits a performance that exceeds that of all previous attempts at electrostatically tunable configurations \cite{maillet2020}. In this study, this significant mechanism was identified only during data analysis, suggesting potential for future refinements to improve the response further.

A critical component of our structural analysis that requires comprehensive characterization is the electron-phonon coupling within the InAs mesas. The elucidation of this interaction requires the assumption that the entire volume of InAs is thoroughly thermalized and that the lattice phonon reservoir is in thermal equilibrium with the cryostat temperature $T_{bath}$.
This coupling follows the usual form \cite{giazottoOpportunities2006}:
\begin{equation}
    \dot{Q}_{e-ph}=\Sigma {\cal V}(T_e^n-T_{bath}^n),
\end{equation}
with $\cal V$ representing the volume of the InAs cold resistor, $\Sigma$ the e-ph coupling constant, and $n$ the exponent of this power law. The specific values of $n$ and $\Sigma$ for InAsOI are shown in Figs. \ref{Fig3} (f) and (g), respectively, as a function of $T_{bath}$. The yellow-shaded area in these panels represents the validity range of our measurements (more information is available in the SI file and in a more detailed study ~\cite{battisti2024}). The values used throughout this work for these two parameters are $n=6.7$ and $\Sigma=5.5\times10^8$ Wm$^3$/K$^n$. 
These values were crucial for designing the entire device architecture and for the numerical simulations presented in the subsequent sections.

\subsection*{Gate-Controlled Photonic Heat Transport Manipulation}
The experiment begins by injecting a current $I_{heat}$ from a floating circuit directly into the hot resistor, thus increasing its electronic temperature $T_{hot}$. The temperature of the cold reservoir $T_{cold}$ also arises due to the radiative heat channel. During this process, we manipulated heat transfer by varying $V_g$, tuning the transport properties of the JoFET (Fig. \ref{Fig1} (b) and Eqs. 1-2) and, consequently, the radiative coupling and the transferred heat flux $\dot{Q}_{\gamma}^{hot \rightarrow cold}$.

\begin{figure*}[t!]
\begin{centering}
  \includegraphics[width=1.1\columnwidth, trim= 0cm 0cm 0cm 0cm]{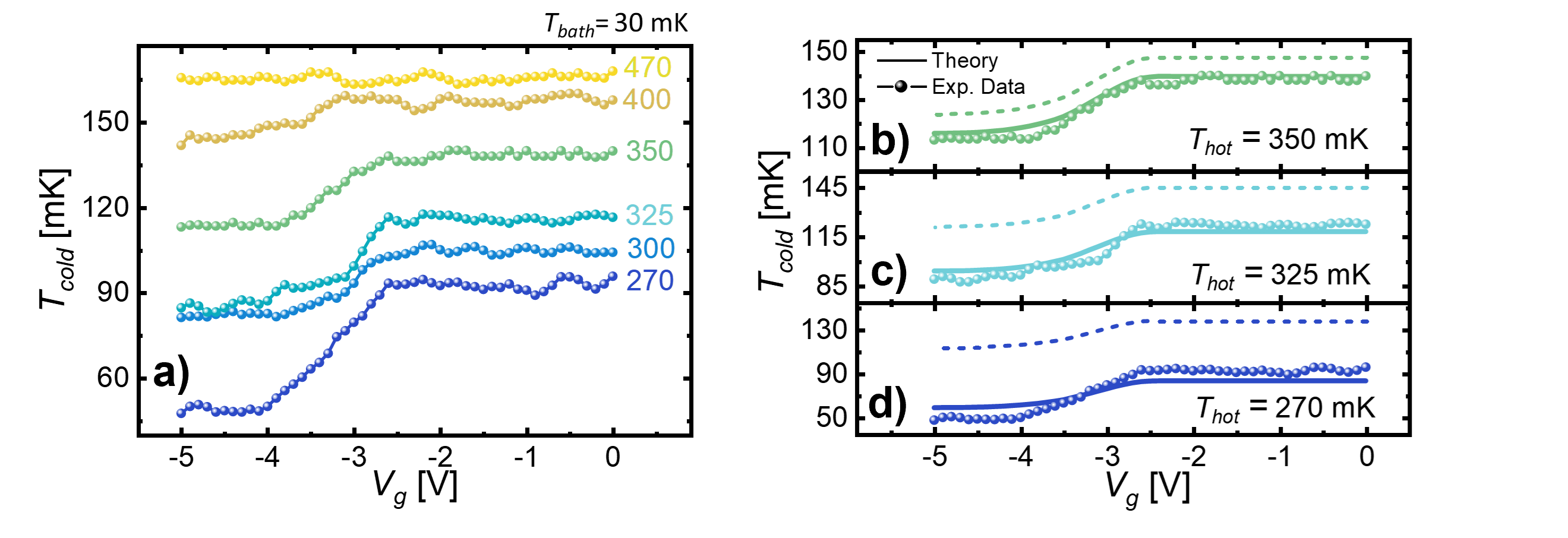}
    \caption{\textbf{Gate-tunable photonic heat transport modulation.} \textbf{a}, Experimental gate-dependent $T_{cold}$ for different $T_{hot}$ (reported in the legend) at $T_{bath}=30$ mK. Each point represents the mean of six switching current measurements taken on the thermometer at the cold reservoir; error bars are smaller than the point size. \textbf{b}-\textbf{d}, Gate-dependent $T_{cold}$ experimental and theoretical data comparison for different $T_{hot}$ at $T_{bath}=30$ mK. Solid lines include the additional loss term $\dot{Q}_{ext}$ while the dashed line does not.}
  \label{Fig4}
\end{centering}
\end{figure*}

Figure \ref{Fig4} (a) shows the results of the radiative heat transport experiment, reporting $T_{cold}$ as a function of the control variable $V_g$ for various $T_{hot}$, keeping the bath temperature fixed at $T_{bath}=30$ mK. For $V_g=0$, we see that higher $T_{hot}$ corresponds to higher $T_{cold}$. This is due to the transmissive photonic heat channel (Fig. \ref{Fig1} (c)), which transfers heat directly from the hot to the cold reservoir. 
Indeed, it is plausible to ignore any direct contribution from the phononic channel as a result of the considerable separation between the two reservoirs.
However, more importantly, when $V_g\leq-2.5$ V, $T_{cold}$ starts to decrease for all the values of $T_{hot}$, except for the highest $T_{hot}=470$ mK. 
This behavior is indeed due to the progressive change in photonic transport resulting from the increase in impedance mismatch induced by $V_g$, which alters the JoFET. 
Indeed, by reducing $V_g$, it first increases its inductance $L_J(V_g)$ and then switches to a resistive state that progressively increases $R_J(V_g)$, as shown schematically by Fig. \ref{Fig3} (c-d). 
This corresponds to a transition to the fully reflective photonic configuration (Fig. \ref{Fig1} (d)), which minimizes photonic heat transport between reservoirs. 

The observation that the curve corresponding to $T_{hot}=400$ mK exhibits minimal influence from $V_g$, while the curve for $T_{hot}=470$ mK remains completely unaffected, can be ascribed to the impact of increasing $T_{hot}$. This increment extends the dominant frequency range determined by the Bose-Einstein distribution, surpassing the tunable operating cut-off of the JoFET, which is typically on the order of a few tens of GHz, due to gate residual capacitance shunting \cite{paghi2024}.

To gain a comprehensive understanding of the behavior of the system for $T_{hot}\leq350$ mK, where substantial modulation occurs, numerical simulations were conducted to retrieve $T_{cold}(V_g)$. Conducting these simulations requires considering that, at $V_g \lesssim V_g^*$, the JoFET transits into a fully resistive state, which may even be represented as a third reservoir within our system. However, as a result of $R_{hot},R_{cold}\geq 4 R_J(V_g)$, it is easy to conclude that the JoFET itself is significantly impedance-mismatched with respect to the other two reservoirs. Consequently, it can be verified that the photon heat current directed from the hot and cold reservoirs toward the JoFET is considerably lower than that between the hot and cold reservoirs, i.e., $\dot{Q}_\gamma^{hot,cold\rightarrow JoFET} \ll \dot{Q}_\gamma^{hot\rightarrow cold}$. 
Considering that the three reservoirs are composed of the same material and exhibit comparable volumes, which consequently results in analogous electron-phonon coupling strengths across all reservoirs, it is ascertainable that the JoFET electronic temperature change is negligible. Thus, it is legitimate to assign its temperature to $T_{bath}$.
Under steady-state conditions, the aggregate of all incoming and outgoing thermal currents within each subsystem of the structure must be identical to $0$, being the overall heat current conserved. Consequently, the temperature of the cold resistor $T_{cold}$ can be determined by numerically solving the following energy balance equation:  
 \begin{equation}
     \begin{split}
    \dot{Q}_\gamma^{hot\rightarrow cold} - \dot{Q}_\gamma^{cold\rightarrow JoFET}
    -\dot{Q}_{e-ph}-\dot{Q}_{ext}=0,
\end{split}
\label{bilancio}
 \end{equation}
where $\dot{Q}_\gamma^{i \rightarrow k}$, dependent on $\tau_{ik}$, is calculated employing the JoFET's parameters reported in Fig. \ref{Fig3} (c,d). For the sake of clarity, we fully include the term $\dot{Q}_\gamma^{cold\rightarrow JoFET}$ in the above energy balance equation, but we have numerically verified that the exclusion of this term would produce a minimal correction less than $1\%$. This substantiates the previously discussed argument, indicating that the JoFET's reservoir can be disregarded even when it is in a dissipative state. Furthermore, in the preceding equation, we also incorporate an additional term $\dot{Q}_{ext}$, which may account for unknown losses; however, we posit that it is independent of the gate (see below).

The numerical calculations are reported in Figs. \ref{Fig4} (b-d), presenting a comprehensive comparison between the experimental data and the results predicted by the model for $T_{cold}(V_g)$, illustrating the model's ability to reproduce the form and dependence on the gate voltage accurately. 
For completeness, we show with solid lines the data obtained from calculations that include the additional gate-independent loss term $\dot{Q}_{ext}$, as opposed to the dashed lines.
Furthermore, the phenomenon underlying the high- and low-temperature plateaus in the $T_{cold}$ vs $V_g$ curves has become more apparent. The initial high-temperature plateau is attributable to the minimal change in $\tau(\omega,V_g)$ resulting from reduced variations in $L_J(V_g)$ at $ V_g > -2.5$V (Fig. \ref{Fig3} (c)), concomitant with the unchanged temperature modulation. Similarly, the low-temperature plateau emerges as a consequence of a flat $R_J(V_g)$ characteristic for $V_g<-4$ V (Fig. \ref{Fig3} (d)), rendering $\tau (\omega,V_g)$ invariant with respect to the specific value of $V_g$ within this domain.

In our calculations, we found that the magnitude of the gate-independent loss term, $\dot{Q}_{ext}$, must be adjusted in response to changes in the hot reservoir temperature to achieve optimal agreement between the simulations (illustrated by solid lines) and the experimental data. Nevertheless, even in the absence of this term, considerable concordance is observed in the dynamic alteration of the cold temperature as a function of the gate voltage, as shown by the dashed lines. Specifically, we estimate values of $\dot{Q}_{ext}=8$, $5$, and $2$ fW for the curves at $270$, $300$, and $350$ mK, respectively. This term accounts for the power dissipation processes inherent in our coupling circuit, which, although currently indeterminate, are deemed unavoidable.

\begin{figure*}[t!]
  \includegraphics[width=1\columnwidth, trim= 0cm 1cm 0cm 0cm]{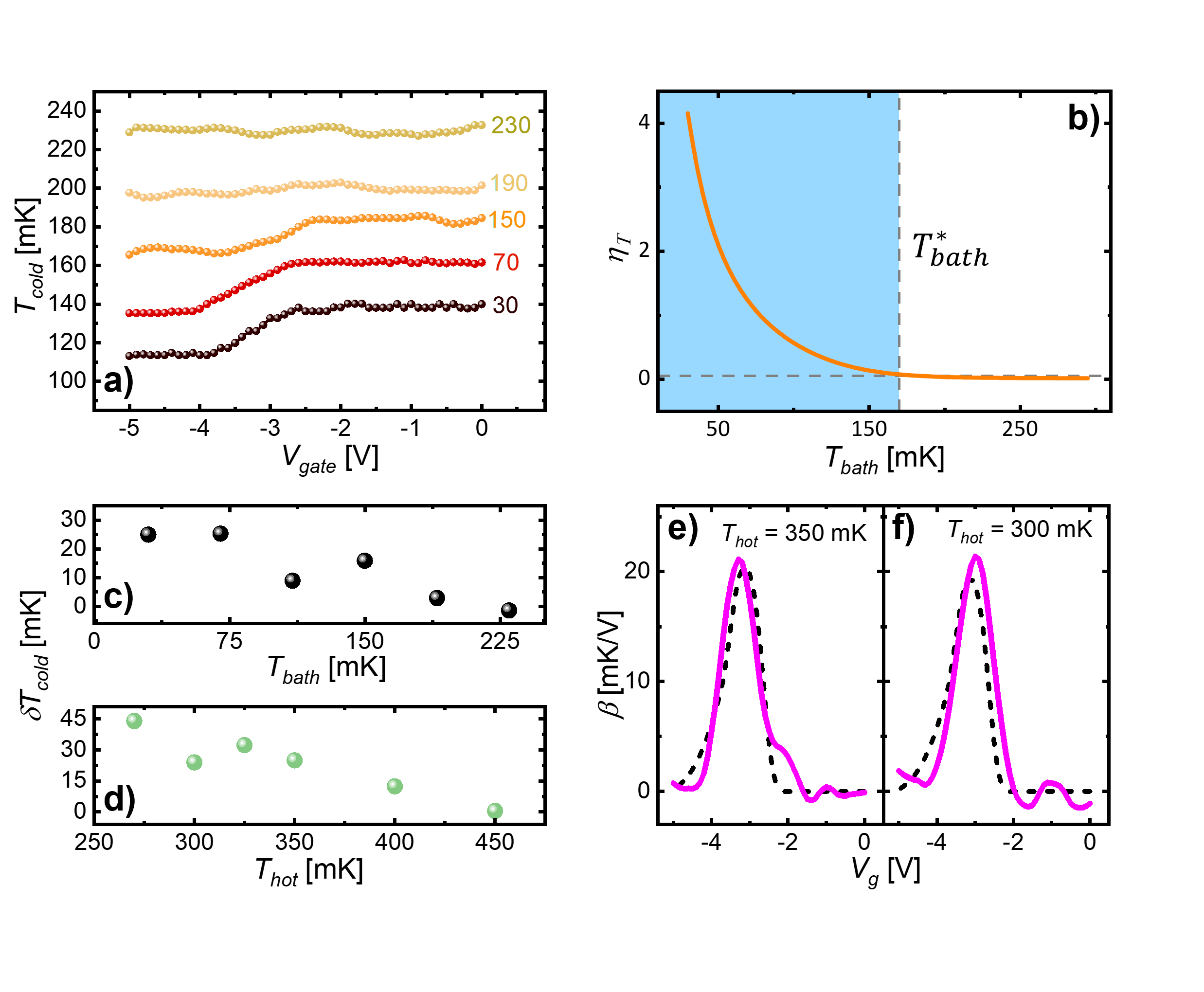}
    \caption{\textbf{Influence of the bath temperature on the gate-tunable photonic heat transport modulation} 
    \textbf{a}, Experimental gate-dependent $T_{cold}$ for different $T_{bath}$ (reported in the legend) at $T_{hot}=350$ mK. Each point represents the mean of six switching current measurements taken on the thermometer at the cold reservoir; error bars are smaller than the point size. \textbf{b}, Numerical simulation of $\eta_T=(T_{cold}-T_{bath})/(T_{bath})$. The horizontal gray dashed line corresponds to $\eta_T^*=0.1$, where the relative difference between $T_{cold}$ and $T_{bath}$ is negligible. The vertical gray dashed line indicates $T_{bath}^*=170$ mK associated with $\eta_T^*=0.1$. The blue-shaded region represents the bath temperature range in which the cold reservoir operates properly. \textbf{c}-\textbf{d}, Temperature difference $\delta T_{cold}$ of the cold reservoir between the JoFET operating in the superconducting state ($V_g>-2V$) and the resistive state ($V_g<-4V$). Black points in panel \textbf{c} report data versus $T_{bath}$ while green points in panel \textbf{d} report data versus $T_{hot}$. \textbf{e}-\textbf{f}, Gate-dependent thermal transimpedance $\beta = \partial T_{cold}/\partial V_g$ for two different selected $T_{hot}$ at $T_{bath}=30$ mK. The magenta curve refers to experimental data, while the black dashed curve indicates theoretical results obtained with the model.}
  \label{Fig5}
\end{figure*}
To assess the robustness of the effect of modulation on photonic transport in relation to bath temperature, measurements analogous to those presented in Fig. \ref{Fig4} (a) were carried out, by which $\dot{Q}_{inj}\simeq2.6$ pW was kept constant while $T_{bath}$ varied. The injected power corresponded to that utilized for the light green curve in Fig. \ref{Fig4} (a) at $T_{hot}= 350$ mK and $T_{bath}= 30$ mK. The results are depicted in Fig. \ref{Fig5} (a), demonstrating that up to $T_{bath}=150$ mK, a modulation in $T_{cold}$ remains discernible; conversely, at higher bath temperatures, the effect disappears due to the augmented e-ph interaction on the InAsOI platform. This finding corroborates our experimental results, theoretically predicted in Fig. \ref{Fig5} (b), when we reported the curve $\eta_T=(T_{cold}-T_{bath})/(T_{bath})$ vs. $T_{bath}$, which denotes the relative difference between $T_{cold}$ and $T_{bath}$ varying bath temperature (a comprehensive discussion of the methodology of this simulation is provided within the Supplementary Information). 
As the photonic heat current $\dot{Q}_{\gamma}^{hot\rightarrow cold}$, at a constant injected power $\dot{Q}_{inj}$, depends on the temperature differential between the reservoirs, an increase in the bath temperature evidently results in a decrease in the relative variation of $T_{cold}$.
For a plausible representative value of $\dot{Q}_{inj}=2.6$ pW, the relative difference between $T_{cold}$ and $T_{bath}$ is not negligible ($\eta_T > 0.1$) only if $T_{bath}<T_{bath}^* = 170$ mK. 
This interval of values of $T_{bath}$, which is expected to be the optimal regime for observing thermal modulation, is highlighted as a blue-shaded area in Fig. \ref{Fig5} (b). For $T_{bath}>T_{bath}^*$, the electron-phonon (e-ph) coupling becomes predominant, thus entirely overshadowing the influence of the photonic heat current $\dot{Q}_{\gamma}^{hot \rightarrow cold}$.

To comprehensively assess the performance of our device, we have derived the maximum temperature variation $\delta T_{cold}$, calculated as the difference between the upper and lower temperature plateaus of the $T_{cold}$ versus $V_g$ curves of Fig. \ref{Fig4} (a) and \ref{Fig5} (a). Figures \ref{Fig5} (c) and (d) report $\delta T_{cold}$ as a function of $T_{bath}$ and $T_{hot}$, respectively. Analyzing $\delta T_{cold}$ against $T_{hot}$, the presence of JoFET allows the modulation of the cold-resistor temperature of $\sim 45$ mK for $T_{hot}=270$ mK, setting the record value of the state of the art ($\gtrsim 20$ times) for the modulation of temperature achieved through electrostatic control \cite{maillet2020}. 
This breakthrough temperature modulation is attributed to the implementation of JoFET on the InAsOI platform, which has been recognized as an optimal candidate for gate-tunable heat transport architectures due to its reduced electron-phonon coupling, compared to traditionally used metals \cite{battisti2024,paghi2025b} and, at the same time, for the better impedance matching determined by the JoFET inductive component.

The $\delta T_{cold}$ vs. $T_{hot}$ shows a monotonic decrease with increasing $T_{hot}$, becoming negligible at $T_{hot}\sim470$ mK, where the effect vanishes. An analogous decrease is observed for $\delta T_{cold}$ vs. $T_{bath}$, reaching negligible levels for $T_{bath} \gtrsim 170$ mK. 
These trends are due to two different phenomena, as explained above, and confirm what was already mentioned. In the first case, an increase in $T_{hot}$ produces a higher-frequency spectrum of thermal radiation, which is more strongly affected by capacitive shunting of the modulating impedance of the JoFET due to gate capacitances. In the latter, the increase of $T_{bath}$ reduces the relative significance of the contribution of the photonic heat current to the definition of $T_{cold}$ (see Fig. \ref{Fig3} (f)).

Figures \ref{Fig5} (e) and (f) illustrate the impact of $V_g$ on $T_{cold}$, calculating the thermal transfer function $\beta = \partial T_{cold}/\partial V_g$ as a function of $V_g$ for $T_{hot}=350$ mK and $T_{hot}=300$ mK, respectively, at $T_{bath}=30$ mK. In particular, for both curves, $V_g$ exerts minimal influence from $-5$ V to $-4.5$ V, then significantly affects $T_{cold}$, achieving values of $20$ mK/V for $V_g \sim V^*$, followed by negligible effects between $-2$ V and $0$ V, in alignment with the properties of the JoFET.
The black-dashed lines depict the results of the theoretical model. The comparative analysis demonstrates that the model accurately predicts gate dependence, as previously described. 
This observation highlights the model's ability to precisely characterize the dynamic behavior of $T_{cold}$ as a function of gate voltage, thereby resolving uncertainties associated with unavoidable gate-independent terms.

\section*{Conclusions}
We have successfully fabricated a hybrid superconducting-semiconducting caloritronic circuit, achieving substantial electrostatic thermal modulation of the cold reservoir up to $\sim 45$ mK in a completely wireless, non-galvanic environment over a notable distance of $\sim 1$ mm. This was accomplished by employing the gating properties of a Josephson field-effect transistor, which enabled modification of the photonic heat transfer from a hot reservoir to a cold reservoir via black-body radiation. Notably, the inductive response of the dissipationless JoFET regime has been shown to exhibit superior impedance matching performance, even compared to simple dissipationless wiring, which compensates for the capacitive response required in non-galvanic setups.

In this context, the demonstration of gate-regulated suppression of supercurrent in \textit{all-metallic} superconducting weak-links \cite{ruf2024gate,paolucci2018ultra,desimoni2019,bours2020} is also identified as a promising alternative mechanism for applications in photonic thermal management.
Our study establishes a new benchmark in the electrostatic manipulation of radiative heat transfer, surpassing the existing literature by more than an order of magnitude \cite{maillet2020}. 

Moreover, the InAsOI platform has been confirmed to be suitable for conducting heat management experiments, owing to its minimal electron-phonon coupling and the straightforward integration of highly sensitive SSmS thermometers. Our device operates at bath temperatures as high as $\sim 170$ mK, marking a significant advancement toward the realization of on-chip heat management and thermal routing systems \cite{giazotto2012,li2012,partanen2016,timossi2018}. These phenomena are of considerable interest in the fields of radiation detection \cite{virtanen2018,singh2025,karimi2024}, energy management within quantum architectures \cite{gubaydullin2022,partanen2018,ibabe2024,nakamura2025,timofeev2009}, quantum thermodynamics tests \cite{pekola2015}, as well as in the advancement of thermal logic gates and memory systems \cite{li2004,wang2007,wang2008}.

\section*{Methods}
\subsection*{Fabrication Methods}
The circuit shown in Fig. \ref{Fig2} (a) was fabricated starting from the InAsOI platform described in the main text. The InAsOI substrate was cleaned in a sonic bath with acetone and isopropanol. First, an electron beam lithography (EBL) step (AR-P 679.04 e-beam resist, dose of $330$ $\mu$C/cm$^2$) with a subsequent thermal evaporation of a $10$/$40$ nm-thick Ti/Au bilayer were performed to obtain alignment markers.

InAs mesas were defined through aligned EBL (ma-N 2403 e-beam resist, dose of $130$ $\mu$C/cm$^2$) and successive wet chemical etching (H$_2$O:C$_6$H$_8$O$_7$:H$_2$O$_2$:H$_3$PO$_4$ $21:5:0.5:0.25$ volumes, with C$_6$H$_8$O$_7$  1M, H$_2$O$_2$ 30\% in mass, and H$_3$PO$_4$ 85\% in mass). After the mesas have been defined, the InAs must undergo surface oxide removal and surface passivation. This process was carried out by leaving the sample in a sulfur supersaturated (NH$_4$)$_2$S$_x$ solution ($0.29$ M of (NH$_4$)$_2$S and $0.3$ M of S in water) at $45^{\circ}$C.

A $120$ nm-thick Al layer was deposited across the entire sample to ensure clean contact with the InAs. Al contacts on the InAs mesas and signal traces were defined through aligned EBL (ma-N 2403 e-beam resist, dose of $130$ $\mu$C/cm$^2$) and selective wet chemical etching of Al (Transene Al etchant type D). 

A 31 nm-thick HfO$_2$ layer was grown by ALD over the entire substrate and used it as a gate, coupling capacitors, and as a trace insulator. Eventually, aligned EBL (AR-P 679.04 e-beam resist, dose of $330$ $\mu$C/cm$^2$) followed by evaporation of a $10$/$120$ nm-thick Ti-Al bilayer were performed to define the top gate control traces and flux lines.

\subsection*{Electrical Measurement Methods}
Electrical measurements were performed in a low-pass filtered (1 kHz cut-off) He$^3$–He$^4$ dry dilution refrigerator (Triton 200, Oxford Instruments) at various bath temperatures ranging from $30$ mK to $830$ mK. The JJs operated as Josephson thermometers (see SI for a comprehensive analysis on thermometer calibrations) were characterized by 4-wire I-V measurements. The readout current was obtained by connecting a voltage source (Yokogawa GS200) in series with a resistance $1000$ times larger than the total measurement setup resistance. The voltage drop across the JJ was amplified at room temperature using a voltage preamplifier (Model 1201, DL Instruments) and measured with a digital multimeter (HP Agilent 34410A). The thermal energy for the hot reservoir was provided by an isolated, battery-operated, floating generator, adjustable via a potentiometer. This voltage generator was also in series with a load resistance $1000$ times higher than the total measurement setup resistance to inject a heating current $I_{heat}$ inside the InAs mesa, as shown in Fig. \ref{Fig2} (a). Measurement of this current was performed using a voltage preamplifier (Model 1201, DL Instruments), which amplifies the voltage drop across the load resistance. The preamplifier output was measured with a digital multimeter (HP Agilent 34410A). The gate voltage for the JoFET operation was provided by a voltage source generator (Yokogawa GS200).

\subsection*{Estimate of $T_{bath}^*$}

The temperature of the cold terminal is determined mainly by two competing heat fluxes: $\dot{Q}^{hot \rightarrow cold}_{\gamma}$ and $\dot{Q}_{e-ph}$, as shown in Eq. \ref{bilancio} of the main text. To give an estimate of the conditions under which one dominates the other, we first neglect $\dot{Q}^{cold \rightarrow JoFET}_{\gamma}$, in the thermal balance of the hot lead since it is much smaller than the other currents involved, as we also numerically verified later (see also the main text). After that, we compute the crossover temperature value $T_{bath}^* \approx 170$ mK,  at which the $V_{g}$-sensitive behavior is expected to fade, using the method presented below. 

We assume injecting a constant heating power of $\dot{Q}_{inj} = 2.6$ pW into the hot reservoir (as experimentally) and compute the thermal balance in this reservoir to calculate $T_{hot}$:
\begin{equation}
    \dot{Q}_{inj} - \dot{Q}_{e-ph}(T_{hot},T_{bath}) = 0.
\end{equation}
In the above equation, we can safely neglect the term $\dot{Q}_\gamma^{hot\rightarrow cold}(T_{cold},T_{bath},V_g)$, as it is roughly three orders of magnitude smaller than the other two. 
Then we can calculate $T_{cold}$ with the simplified thermal balance in the cold reservoir:
\begin{equation}
    \dot{Q}_\gamma^{hot\rightarrow cold}(T_{hot},T_{cold},T_{bath}) - \dot{Q}_{e-ph}(T_{cold},T_{bath})=0\ .
\end{equation}
where we assume that $V_g$ is fixed so that the photonic heat channel is only transmissive.
Then we can calculate the relative difference between $T_{cold}$ and $T_{bath}$ which can be expressed in terms of the relative variation $\eta_T=(T_{cold}-T_{bath})/(T_{bath})$ to evaluate the impact of photonic transport with respect to the e-ph coupling in the cold reservoir. The results of this relative difference as a function of $T_{bath}$ are shown in Fig. \ref{Fig5} (b) of the main text.
This figure clearly illustrates the expected behavior, where $T_{cold}$ approaches $T_{bath}$ more closely as $T_{bath}$ increases. To specifically define a crossover temperature, we identify the point at which the relative variation reaches $10\%$ ($\eta_T^{*}=0.1$), as indicated by the horizontal gray dashed line in the accompanying figure. The value obtained is indeed $T_{bath}^*=170$ mK. Consequently, the blue-hued region in the figure delineates the experimentally relevant range of $T_{bath}$, where the influence of $T_{cold}$, attributed to $\dot{Q}^{hot \rightarrow cold}_{\gamma}$, is significant. Within this interval, the modulating effect of $V_{gate}$ is expected to be detected.

\section*{Aknowledgements}
SB, MP, GdS, AB, AP, and FG acknowledge the PNRR MUR project PE0000023-NQSTI for partial financial support. 
AB acknowledges MUR-PRIN 2022 - Grant No. 2022B9P8LN-(PE3)-Project NEThEQS “Non-equilibrium coherent thermal effects in quantum systems” in PNRR Mission 4-Component 2-Investment 1.1 “Fondo per il Programma Nazionale di Ricerca e Progetti di Rilevante Interesse Nazionale (PRIN)” funded by the European Union-Next Generation EU and CNR project QTHERMONANO. 
LS and GS acknowledge the National Infrastructure of the National Research Council (CNR) for quantum simulations and calculations, “PASQUA."

\printbibliography

\appendix

\setcounter{equation}{0}
\setcounter{figure}{0}
\setcounter{page}{1}
\renewcommand{\theequation}{S\arabic{equation}}
\renewcommand{\thefigure}{S\arabic{figure}}

\section*{\Large Supplementary Information}

\begin{figure*}[t!]
  \includegraphics[width=1\columnwidth, trim= 0cm 1cm 0cm 0cm]{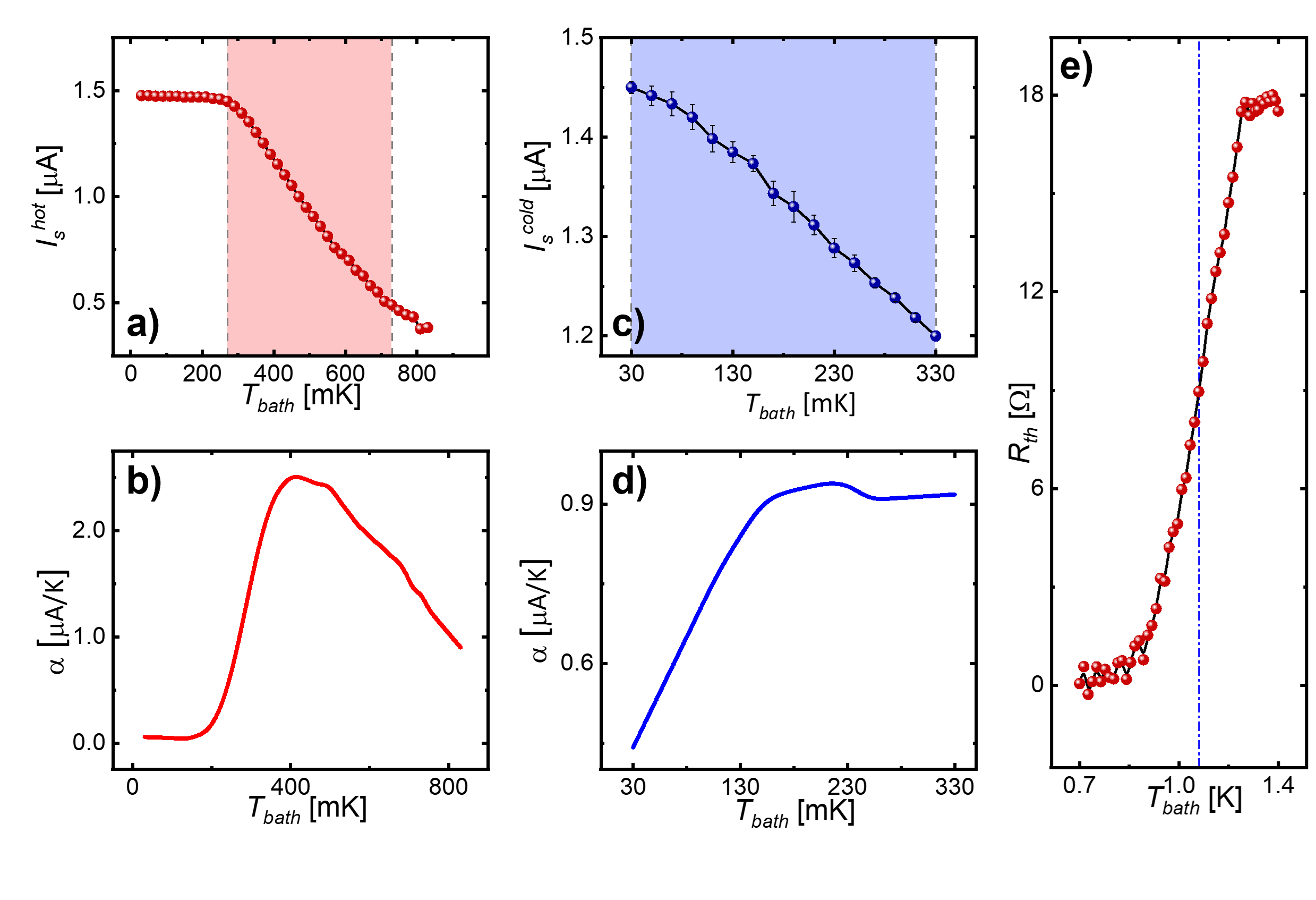}
    \caption{\textbf{Calibration of the thermometers and switching temperature.} \textbf{a}, Switching current $I_s^{hot}$ and \textbf{b} sensitivity $\alpha=\partial I_s^{hot}/\partial T_{bath}$ as a function of $T_{bath}$ of the thermometer on the hot reservoir.
    The experimental data and error bars (when not visible, they are smaller than the point size) represent the mean and standard deviation over six values of the switching current. The shaded region indicates that this thermometer remains operational, i.e., $\alpha>0$, over the range from 260 mK to 730 mK. The solid curve is the result of a polynomial interpolation, which serves as the calibration curve to infer the electronic temperature $T_{hot}$ when a switching current $I_s^{hot}$ is measured. \textbf{c} Switching current $I_s^{cold}$ and \textbf{d} sensitivity $\alpha=\partial I_s^{cold}/\partial T_{bath}$ of the thermometer on the cold reservoir as a function of $T_{bath}$. The statistical analysis of the data and their representation are identical to panels (a) and (b). 
    The shaded region indicates that this thermometer is usable only in the temperature range of 30 mK to 330 mK. From this calibration curve, one can determine the $T_{cold}$ from the measurement of $I_s^{cold}$. \textbf{e}, Resistance of the JJ on the hot reservoir $R_{JJ}$ as a function of $T_{bath}$. This panel shows the transition from the normal to the superconducting state, which occurs at the switching temperature $T_c\simeq 1.12$K. Each point and error bar (not visible when smaller than the point size) in this graph is the result of the mean and standard deviation of 5 resistance measurements.}
  \label{FigS2}
\end{figure*}

\section*{Calibration of the Josephson Thermometers}

Josephson thermometers were realised using JJs measured with a 4-wire measurement setup (see Fig. 2(a) in the main text). A readout current $I^{th}_{hot,cold}$ is injected, while the voltage drop across the JJ $V^{th}_{hot,cold}$ is collected. From the I-V curve, the switching current $I_s^{hot,cold}$ was obtained. The bath temperature $T_{bath}$ was adjusted to retrieve the $I_s$ vs $T_{bath}$ relationship for each JJ and used as a calibration file for the Josephson thermometers. The $I_s(T_{bath})$ was used to infer the electronic temperature of the InAs mesas of the hot reservoir heated by the injected current $I_{heat}$ and the cold reservoir heated by radiative thermal transport. The calibration measurements for the thermometers of the hot and cold reservoirs are shown in Fig. \ref{FigS2} (a,c). The shaded region indicates the operational regime of each thermometer, showing that the detectable temperature for the cold reservoir ranges from 30 mK to 330 mK. There was no need to calibrate that thermometer for higher temperatures, since the experiment's operating range for the cold reservoir did not exceed that. For the hot reservoir, the thermometer's operating range is between 260 mK and 730 mK. 
Those ranges correspond to non-negligible values of the sensitivity $\alpha=\partial I_s/\partial T_{bath}$, as shown in Fig. \ref{FigS2} (b,d).\\
The two Josephson thermometers have different operational ranges as a result of the different lengths of their JJs. The one in the hot reservoir is $100$ nm shorter than the one in the cold reservoir, resulting in a higher sensitivity to higher temperatures for the JJ in the hot reservoir.

The critical temperature $T_c$ of the JJ in the hot reservoir was also measured; the results are shown in Fig. \ref{FigS2} (e). A $T_c=1.12$K was obtained, in agreement with the precedent values achieved on the Al-proximitized InAsOI platform \cite{paghi2025c,paghi2025b}.

\section*{Electrical Characterization of the twin JoFET}

\begin{figure}[t!]
\begin{centering}
  \includegraphics[width=0.9\columnwidth, trim= 0cm 0cm 0cm 0cm]{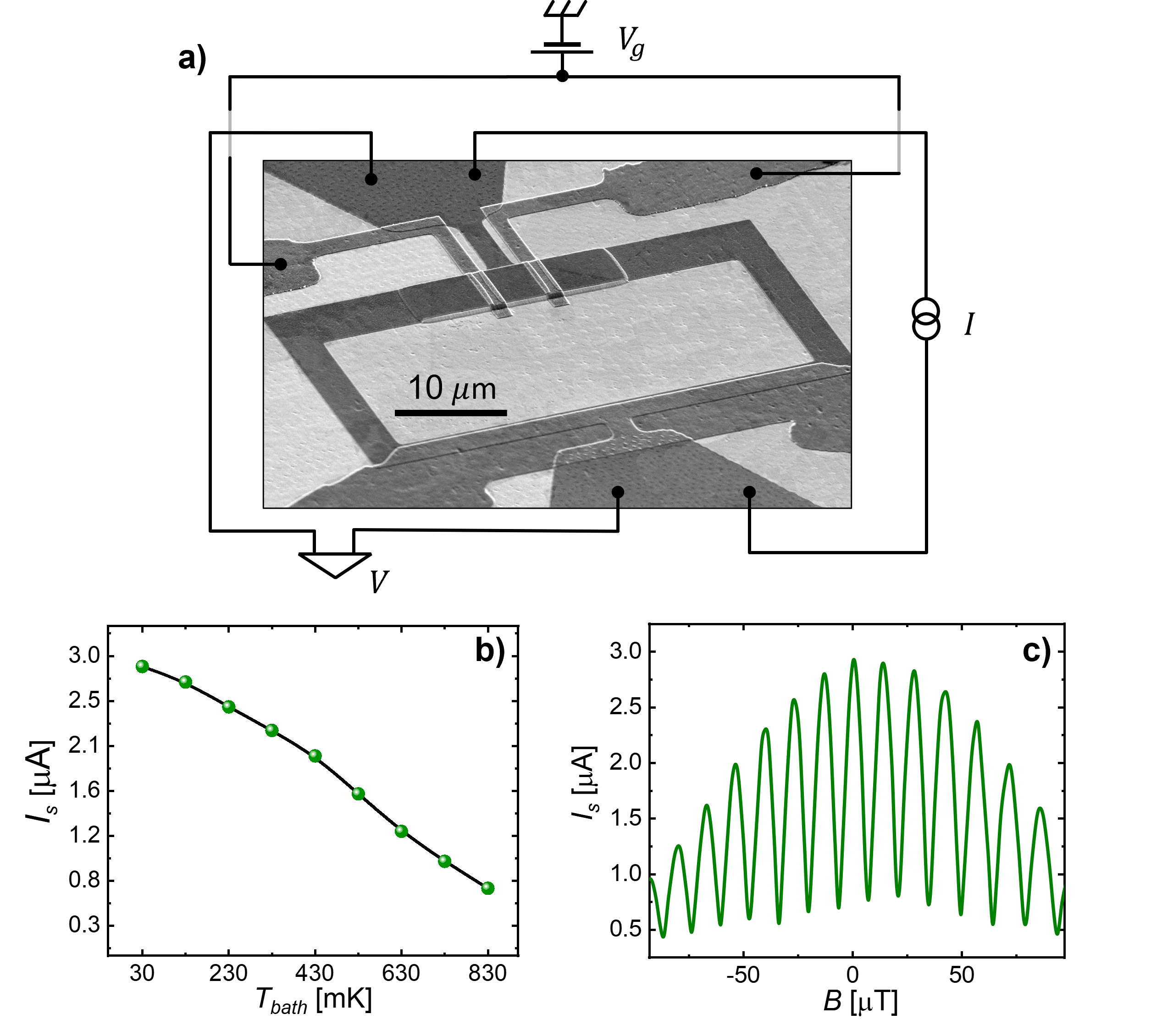}
    \caption{\textbf{JoFET twin architecture present on the same chip.} \textbf{a}, SEM image of the twin JoFET architecture with the wiring setup used for the measurements. The two JoFETs are also embedded in a SQUID loop, as in the coupling circuit of the studied device. The flux line that passes over the loop is also visible, but it was not used in the experiment. \textbf{b}, Switching current $I_s$ of the twin JoFET architecture as a function of the bath temperature $T_{bath}$. \textbf{c}, Switching current of the twin JoFET architecture as a function of the out-of-plane magnetic field $B$.
    }
  \label{FigS3}
\end{centering}
\end{figure}
The electrical performance of the JoFET architecture was evaluated on a twin architecture fabricated on the same chip (Fig. \ref{FigS3} (a)), which was useful to avoid the impacts of parasitic elements due to the fabrication of the bonding pads and the relative measurement setup.

Fig. \ref{FigS3} (b) shows the dependence of $I_s$ on $T_{bath}$, where a monotonic decrease trend is observed as expected. Figure \ref{FigS3} (c) reports the dependence of $I_s$ as a function of the out-of-the-plane magnetic field $B$, where the $B$-tuned typical SQUID-like behavior is apparent. In particular, the interference pattern minima show $I_s$ of about $\sim 500$ nA, instead of reaching a null switching current. This results in reduced variation of the $B$-tuned Josephson inductance $L_J$, causing a negligible impact on the modulation of the photonic heat current $\dot{Q}_\gamma ^{hot \rightarrow cold}$. The results achieved ruled out the use of SQUID-like interference as an effective method for tuning photonic heat transport in our experimental configuration.

\section*{Characterization of the Electron-Phonon Coupling on the InAsOI Platform}
\begin{figure}[t!]
\begin{centering}
  \includegraphics[width=0.8\columnwidth, trim= 0cm 1.3cm 3cm 0cm]{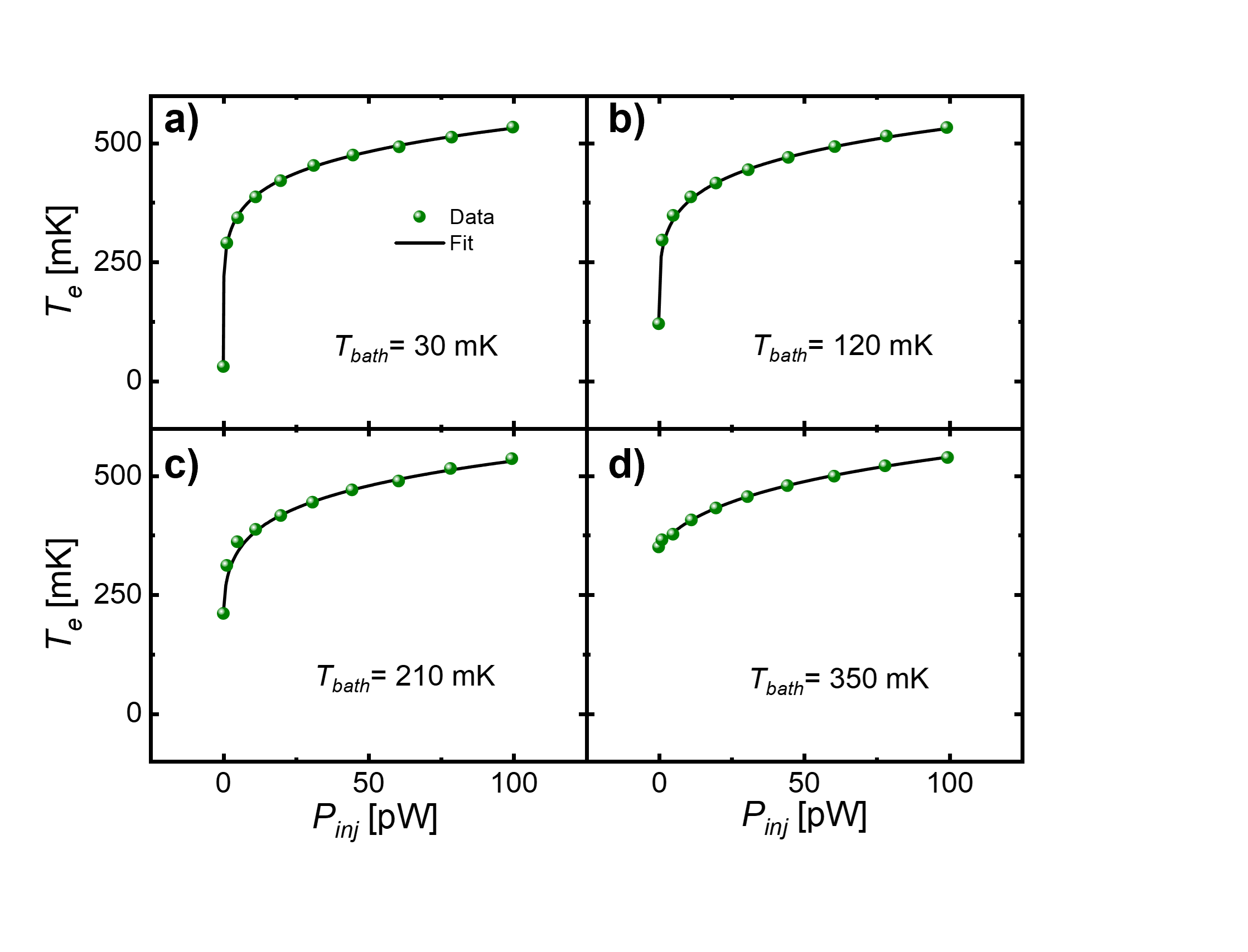}
    \caption{\textbf{a}-\textbf{d}, These panels show preliminary measurements of e-ph coupling on this heterostructure. In particular, the electronic temperature $T_e$ was measured as a function of the injected power $P_{inj}$ for different $T_{bath}$ to extract the exponent $n$ and the coupling constant $\Sigma$ of the e-ph coupling formula. Experimental data with the best fits are shown for four selected $T_{bath}$ values to demonstrate optimal agreement.}
  \label{FigS4}
\end{centering}
\end{figure}
To determine the electron-phonon (e-ph) interaction parameters for accurate simulations and optimal sample design, a comprehensive characterization of the e-ph coupling within the specific heterostructure was conducted.
The setup to perform these measurements is the one shown in Ref. \cite{battisti2024}, where a power $P_{inj}$ is injected into an InAs mesa of volume $V$ through Joule heaters and supposing that the electrons can only exchange heat with the lattice phonons, the thermal balance can be written as $P_{inj}-\Sigma V(T_{e}^n-T_{bath}^n)=0$, where the electronic temperature $T_e$ can be extracted as a function of $P_{inj}$ in a fixed $T_{bath}$ with $\Sigma$ and $n$ as fitting parameters:
\begin{equation}
    T_e=\sqrt[n]{\frac{P_{inj}}{\Sigma V}+T_{bath}^n} \vspace{3mm}.
    \label{T_e}
\end{equation}
Indeed, the electronic temperature $T_e$ can be measured as a function of $P_{inj}$ using a thermometer on the InAs mesa, and fitting the $T_e(P_{inj})$ curves with Eq. \ref{T_e}, one can find the fitting parameters $\sigma$ and $n$. 
The fittings are shown in Fig. \ref{FigS4} (a,b,c,d), where four selected $T_e$ vs $P_{inj}$ curves for four different $T_{bath}$ are displayed to see the optimal accordance with the experimental data. The results for $n$ and $\Sigma$ are shown in Fig. 3 (f,g) of the main text, where the points and error bars are the result of a best fit procedure with Eq. \ref{T_e}. We observe that for $T_{bath}<255$ mK (the yellow shaded region) the fitting values are substantially constant. We think that outside that region the fittings lose significance. We estimated that this is due to inhomogeneous thermalization of the InAs mesa at higher bath temperatures. Indeed, for such temperatures $T_{bath}\sim 250$ mK, the e-ph relaxation length $\lambda_{e-ph}$ becomes comparable to the mesa dimension \cite{battisti2024}. 

\end{document}